\documentclass[lettersize,journal]{IEEEtran}
\usepackage{amsmath,amsfonts}
\usepackage{algorithm}
\usepackage{array}
\usepackage[caption=false,font=normalsize,labelfont=sf,textfont=sf]{subfig}
\usepackage{textcomp}
\usepackage{stfloats}
\usepackage{url}
\usepackage{enumerate}
\usepackage{verbatim}
\usepackage{graphicx}
\usepackage{xcolor}
\usepackage{cite}
\usepackage{multirow}
\usepackage{algorithm}
\usepackage{algpseudocode} 
\usepackage{tabularx,booktabs,textcomp}
\usepackage{amsthm}

\usepackage{amsmath,amssymb,amsthm}

\begin{document}
% \title{Adversarial Attacks on EV Charging Stations: FGSM-Based Anomaly Generation and LSTM-Based Detection}
% \title{Enhancing Cyber Security of Electric Vehicle Charging Infrastructure through Anomaly Detection and Spoofing Classification}
%\title{FGSM based adversarial attacks on EV charging stations and anomaly detection using LSTM-based autoencoder}
% \title{LSTM Autoencoder based Anomaly Detection of Adversarial Cyber Attacks generated through FGSM for EV Charging Stations}
% \title{TCN-AE based anomaly detection for multi-replay attacks in EV charging stations}
% \title{Energy anomaly detection for multi-replay attacks on EV charging stations}
\title{Methodology for Detecting Energy Anomalies due to Multi-Replay Attacks on Electric Vehicle Charging Infrastructure}
\author{{Sagar Babu Mitikiri$^{a}$, Vedantham Lakshmi Srinivas$^{b}$, Mayukha Pal$^{*c}$}
 %~\IEEEmembership{Member,~IEEE}
 %~\IEEEmembership{Senior Member,~IEEE}
        % <-this % stops a space
\thanks{(Corresponding author: $^{*}$Mayukha Pal)}
\thanks{$^{a}$Sagar Babu Mitikiri is a Data Science Research Intern at ABB Ability Innovation Center, Hyderabad 500084, India and also a Ph.D. Research Scholar at the Department of Electrical Engineering, IIT (ISM) Dhanbad, Dhanbad, 826004, India.}
\thanks{$^{b}$Dr. Vedantham Lakshmi Srinivas is an Asst. Professor in the Department of Electrical Engineering, IIT (ISM) Dhanbad, Dhanbad, 826004, India.}
\thanks{$^{c}$Dr. Mayukha Pal is with ABB Ability Innovation Center, Hyderabad-500084, IN, working as Global R\&D Leader – Cloud \& Advance Analytics (e-mail: mayukha.pal@in.abb.com).}}

% The paper headers
%\markboth{Journal of \LaTeX\ Class Files,~Vol.~, No.~, }%
%{Shell \MakeLowercase{\textit{et al.}}: A Sample Article Using IEEEtran.cls for IEEE Journals}

\maketitle
\begin{abstract}

The increasing deployment of Electric Vehicle Charging Infrastructures (EVCIs) introduces cybersecurity challenges, particularly due to inherent vulnerabilities, making them susceptible to cyberattacks. The vulnerable points in EVCI are charging ports, which serve as the links between the EVs and the EVCI as they transfer the data along with the power. Data spoofing attacks targeting these ports can compromise security, reliability, and overall system performance by introducing anomalies in operational data. An efficient method for identifying the charging port current magnitude variations is presented in this research. The MATLAB/SIMULINK environment simulates an EVCI system for various data generating scenarios. A Temporal Convolution Network - Autoencoder (TCN-AE) model is used in training the multivariate time series data of EVCI and reconstructing it. The abnormalities in data are that various charging port current magnitudes are replaced with their respective data of different durations, thus enabling the replay attack scenarios. To detect anomalies, the error between the original and reconstructed data is computed, and these error values are used for detecting the anomalies. With the help of the mean vector and covariance matrices of the errors, the anomaly score is computed in the form of Mahalanobis distance. The threshold is obtained from the short sub-sequence of the errors and optimized for the whole time series data. The obtained optimal threshold is compared with the anomaly score to detect the anomaly. The model demonstrates robust performance in data reconstruction by identifying anomalies with an accuracy of 99.64\%, to enhance the reliability and security in operations of EVCI.

\end{abstract}

\begin{IEEEkeywords}
EVCI, spoofing, replay attacks, anomaly detection, TCN-AE. 
\end{IEEEkeywords}

\section{Introduction}
\label{section: Introduction}
The transportation sector contributes significantly to greenhouse gas emissions, as it uses 62$\%$ of the global oil resources \cite{solaymani2019co2}. Electrifying the transport industry is an important solution and a reflecting trend for decarbonization to reduce the emission of CO\textsubscript{2} at least 3\% annually \cite{CO2emission}. The electrification process includes the substitution of electric vehicles for internal combustion engine vehicles (ICEV). The market sales of EVs exhibited robust growth with 14.1 million units sold in 2023, which signifies a 35$\%$  increase in sales as compared to 2022 \cite{EVsales}. Given that EVs take longer to charge and that the number of EVs is growing, EVCI development is crucial to playing a function akin to that of ICEV gas charging stations \cite{chakraborty2024planning}.

The advancement of EVCIs requires continuous adaptation to support the growing fleet of EVs. This adaptation involves scheduled maintenance, economic optimization, integrating fast charging, and demand response management. Achieving these functionalities necessitates compliance with standardized power and communication protocols \cite{das2020electric}.

As critical access points to the energy infrastructure, charging ports in EVCIs are susceptible to cyber threats \cite{hamdare2023cybersecurity, industrydigits}. Malicious actors exploit these vulnerabilities to launch cyberattacks, leading to anomalies in EVCI data streams. Various types of cyber threats targeting EVCIs include Denial of Charging (DoC), malware infiltration (e.g., ransomware attacks), energy theft, false data injection, and spoofing \cite{acharya2020cybersecurity}. Among these, spoofing attacks are particularly prevalent, causing disruptions in EVCI data integrity. These attacks can occur across different system layers, necessitating focused anomaly detection techniques for charging port data monitoring.

The primary interface that enables communication between EV and EVCI is the charging port, which is a vulnerable and critical component \cite{nasr2022power, ahmed2016electric}. Tampering with charging current magnitudes can pose significant risks including battery safety hazards, financial losses, reputational damage, and grid instability. Adversaries take advantage of the vulnerabilities in the infrastructure and create false demand in the EVCI, by manipulating the charging current values and replacing the duration of the noncharging periods with the charging periods and vice versa (replaying the data scenarios). Replay attacks are the most common type of attacks \cite{elsaeidy2021hybrid, ramanan2021blockchain, taher2023analyzing}, where the adversaries don't need to know the physics of the system. These attacks are formulated by relaying on the past data values at present durations. These replay attacks \cite{le2024comprehensive} deceive the sensors and relays as their values are similar to the normal operation values. Since these attacks fail to damage the potential of the charging systems, they may affect the integrity and repudiation of the charging infrastructure and also cause financial loss to both EV owners and the charging aggregators \cite{humayed2017cyber, acharya2020cybersecurity, van2019non}. Therefore, this work primarily focuses on anomaly detection of the charging port current magnitudes that arise due to replay attacks within EVCI.

\subsection{Related works}
% \label{Section: related work}

Detecting anomalies in EVCI data is essential for preventing potential damage, maintaining reliability, and safety, as well as cybersecurity. Malicious entities can exploit system vulnerabilities to launch cyberattacks, disrupting normal EVCI operations. These disruptions may lead to service outages, infrastructure or vehicle damage, and financial losses. To address these challenges, researchers have proposed various cybersecurity measures to enhance the resilience of EVCI systems \cite{acharya2020cybersecurity, hamdare2023cybersecurity, chandwani2020cybersecurity, mitikiri2023modelling, farhadi2022charging}.

The comparison between the Resnet autoencoder and Physics based models for anomaly detection in charging stations is done in \cite{mavikumbure2023physical} where it is found that the Resnet autoencoder model surpasses in terms of F1 score and accuracy, with benefits like unsupervised learning and nonlinear automated features. Hidden Markov models based defense strategy is developed to identify the potential threats, and a STRIDE based threat modelling is also implemented in \cite{girdhar2021hidden}. The anomalies related to power supply systems of intelligent charging station are detected through Multi Head Attention (MHA) models whose data correlations are generated by EVCI traffic \cite{li2020detecting}.

Extensive research has been conducted on detecting anomalies in EVs during the charging process \cite{mitikiri2025anomaly}. Depending on the lithium power supply operating principles in the EV charging process, the battery failure causes are diagnosed through fuzzy mathematical theory \cite{6023193}. A multivariate Gaussian distribution model is proposed in \cite{yang2022electric} for detecting anomalies in the EV charging data to improve the charging process safety.

DoS attacks are detected by using a deep learning based intrusion detection scheme (IDS) \cite{basnet2020deep} for detecting Denial-of-Service (DoS) attacks. The comparison of Deep Neural Network (DNN) and Long Short-Term Memory (LSTM)-based approaches is done, with results indicating that LSTM outperforms DNN across the evaluated metrics. The authors of the \cite{guo2020cyberattack} detect the cyberattacks in the EVs using a physics-guided machine learning based cyber attacks detection for EVs to differentiate between the physical faults and the cyber attacks on the EV motor drives. 

A regression based model is used for charging data anomaly detection to prevent energy theft with high accuracy, but focuses mainly on state of charge ($SoC$) variations of EVs \cite{babu2024regression}. Similar data-driven analysis is used in \cite{wang2021data} to diagnose problems in charging capacity of the battery. A tree-based model that predicts battery charging capacity over time and considers the battery's SoC and average temperature is utilized.

\begin{figure*}[]
    \centering
    \includegraphics[scale=0.2]{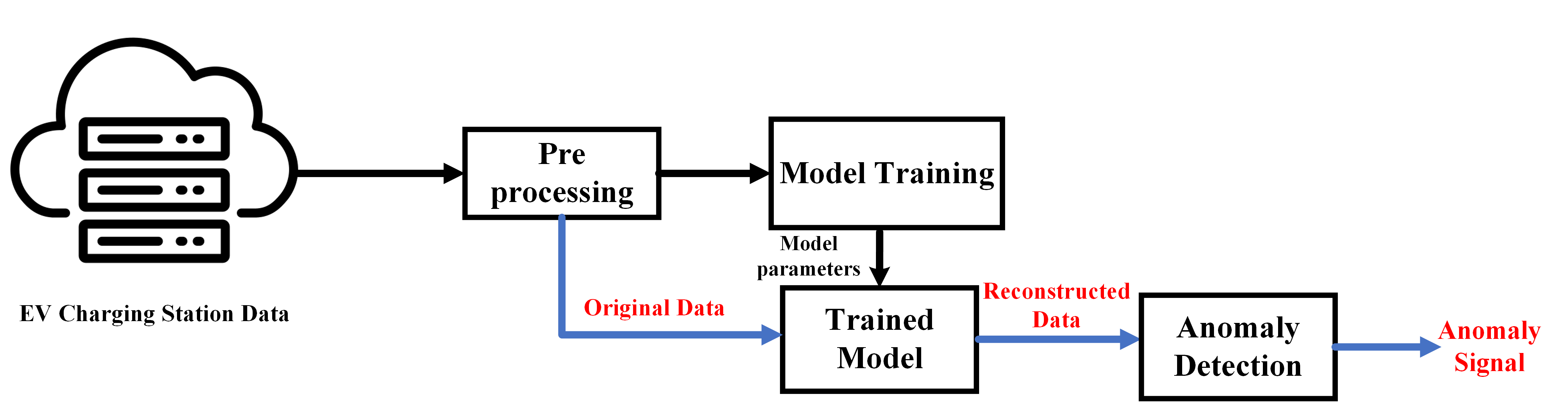}
    \caption{Workflow for the proposed methodology: Black line indicates the process of model training and blue line indicates the anomaly detection process}
    \label{fig: workflow}
\end{figure*}

\subsection{Contributions and Paper Organization}
% \label{Section: Contributions and Paper Organization}

A methodology for anomaly detection of EVCI current values is presented in this article that could be implemented in real time charging stations. A DC EVCI is simulated, as shown in Fig. \ref{fig: EVCI} to validate the proposed methodology. The proposed model is trained with the obtained data to minimize the reconstruction error, and then the anomaly detection methodology is applied by sliding a window of fixed length over the reconstruction error. The following are the main contributions offered in this work:
\begin{itemize}
    % \item A grid-connected DC EV charging station (EVCS) is simulated, and the data is obtained for various scenarios. as shown in Fig. \ref{fig: EVCI}.
    % \item A novel workflow for generating and detecting te anomalies in the EV charging ports current data is proposed as shown in Fig. \ref{fig: workflow}.
    \item Detection of multi-replay attacks in EVCI is presented for the first.  
    \item The TCN-based autoencoder is proposed for training and testing the multivariate time series data from the EVCI.
    \item Anomaly detection is performed using Mahalanobis distance by computing the error between original and reconstructed time series data through a sliding window.
    \item The proposed methodology is applied to an EVCI with multiple charging ports and the model performance demonstrates it's applicability to real world scenarios of data from multiple charging ports of an EVCI.
     
     % demonstrates robust anomaly detection capabilities in scenarios where multiple data variables are manipulated.
     % The anomaly score for the reconstructed multivariate time series data is computed in the form of a Mahalanobis distance for the error between the reconstructed and the original time series data though a sliding window.  
    % The acquired data is trained with a TCN-based autoencoder to learn the under lying dynamics and deep neural architecture is used for obtaining the reconstructed data.
    % \item The anomaly threshold is obtained by optimizing the local threshold value for a short subsequence data.
    % \item The proposed methodology demonstrates robust anomaly detection capabilities in scenarios where multiple data variables are manipulated.
\end{itemize}

This paper starts with introducing the EVCI and highlighting its importance and inherent vulnerabilities in Section \ref{section: Introduction}.  It then provides an overview of the current anomaly detection techniques employed by the EVCI and authors' contributions. Section \ref{Section: TCN-AE and Anomaly Detection} describes the preliminaries about the temporal convolutional networks and their autoencoder derivatives. It also explains the anomaly detection procedure using TCN-AE. Section \ref{Section: System Modelling and Data Acquisition} describes the simulated EVCI architecture and the various data parameters obtained from it. It also explains the preprocessing of the acquired data parameters, and the application of methodology for anomaly detection is elaborated. In Section \ref{Section: Results and Discussions}, the TCN-AE model's performance is evaluated and compared with a few other models. It also provides the visualization of various charging port current magnitudes, highlighting their anomalies. The efficacy of the anomaly detection methodology is also evaluated with the metrics. Finally, the summary of research findings and the achievements are provided in Section \ref{Section: Conclusion} along with the potential areas for future work.

\section{Materials and Methods}
\label{Section: TCN-AE and Anomaly Detection}

\subsection{Temporal Convolution Network - Auatoencoder}

Recurrent architectures are the most common approaches of deep learning practitioners for sequence modeling. These recurrent architectures (also known as recurrent neural networks (RNN)), are capable of capturing the temporal dependencies in the data. Unlike the feed-forward neural networks, which analyze one input at a time independently, RNNs operate on loops, allowing the information to persist over time and making them appropriate for applications such as time series prediction, natural language processing, and speech recognition. Despite capturing the temporal dependencies, RNNs do possess some limitations, like exploding and vanishing gradients and difficulties in capturing long-term dependencies. 

One type of deep neural network that is particularly useful for processing and analyzing visual data is called a convolutional Neural Network (CNN). The properties of CNN, such as translational variance, weight sharing, and computational efficiency help them outperform several other models in various visual tasks including image segmentation, object detection, and image classification. These properties of CNNs make them capable of processing univariate and multivariate time series data when employed with 1D convolutions.

A relatively recent class of neural networks called TCNs were developed to analyze sequential data like time series and natural language processing tasks by combining the principles of CNNs and RNNs. This article uses the TCN-AE for detecting the anomalies in the sequential time series data.  that has fewer dense weights than the other.

\subsubsection{Dilated Convolutions}
The digital filters present in the convolutional layers of the neural networks help in eliminating or amplifying the specific components of the signal. This filtration operation is termed as convolution. In the context of one dimensional signal $x[n]$, where $x[n]$ represents the $n^{th}$ element of the signal and $x$: $\mathbb{T} \rightarrow \mathbb{R}$ with $\mathbb{T} = \{ 0,1,...,T-1 \}$, the convolution of $x[n]$ with a finite impulse response filter $h[n]$, where $h: \{ 0,1,...,k-1 \} \rightarrow \mathbb{R}$, is usually represented as:

\begin{equation}
    y[n] = (x * h)[n] = \sum_{j=0}^{k-1} h[j] \cdot x[n-j]
\end{equation}

where the output of convolution operation is represented by $y[n] \in \mathbb{R}$, the weight of the filter is denoted by $h[i] \in \mathbb{R}$ for the length being $k$, and $*$ is the convolutional operator. The convolution operator is performed by sliding a filter window of length $k$, with $h[i]$ weights, over the sequences of input $x[n]$, and calculating the weighted averages of $x[n]$ with the same weights for each time step which results in a one-dimensional output signal of length $1+T-k$. To ensure the length of the output signal is the same as the input signal, zero paddings are applied to the input sequence initially before applying the filter. This process is commonly known as one-dimensional convolution because the sliding operation by the filter is done only along the temporal axis. The behavior of filter (high pass or low pass) is determined by the $h[n]$. The fundamental concept of CNN involves learning the appropriate weights of filter based on the specific learning task, rather than predetermining $h[n]$.

Consider multivariate time series signal $\textbf{x}[n]$ with a dimensionality of $d$, denoted as $\textbf{x}: \mathbb{T} \rightarrow \mathbb{R}^d$. When treated with convolutional layers, each dimension of $\textbf{x}_j[n]$ is individually convoluted with its corresponding sub-filter $\textbf{h}_j[n]$, where $\textbf{h}: \{ 0,1,....,k-1 \} \rightarrow \mathbb{R}^d$. The resulting output $y[n]$ which is one-dimensional is expressed as a dot product which is given as follows:

\begin{equation}
    y[n] = (\textbf{x}*\textbf{h})[n] = \sum_{j=0}^{k-1}\textbf{h}[j]^T \textbf{x}[n-j]
\end{equation}

Unlike the above standard convolution method, the dilation convolution contains an extra parameter called the dilation rate $q \in N$. It specifies the number of bypassed elements of $\textbf{x}[n]$ between the consecutive filter terminals $h[i]$ and $h[i+1]$, and this dilated operation is represented as:
\begin{equation}
    y[n] = (\textbf{x}*_q\textbf{h})[n] = \sum_{j=0}^{k-1}\textbf{h}[i]^T \textbf{x}[n-qj]
    \label{eq:dilated convolution}
\end{equation}

For $q=1$, the operation will be the same as the original convolution. Using larger dilation results in top-level output that reflects a greater range of inputs, by effectively expanding the receptive field of the convolution networks \cite{archive}. In the eq. \ref{eq:dilated convolution}, if the future values of the input sequences $x[n]$ are processed to produce the output $y[n]$, then this type of convolution are termed as acausal convolution and is given by:
\begin{equation}
    y[n] = (\textbf{x}*_q\textbf{h})[n] = \sum_{j=0}^{k-1}\textbf{h}[i]^T \textbf{x} \left[n-q \left( j - \frac{k}{2} \right)\right]
    % \label{eq:dilated convolution}
\end{equation}

Since the above equations described compute the convolution only for the one-dimensional signal $y[n]$ (output), the practical convolution layers typically comprise numerous discrete filters, and these obtained individual outputs are stacked together which is called as feature map. When a convolution layer of $n_{filters}$ is applied to an input signal $\textbf{x}[n]$ of $T_{train}$ length, the resulting feature map has the dimension $T_{train} \times n_{filters}$ after padding. Each filter's weights, $\textbf{h}[i]$, are regarded as learnable parameters that are often learned using back-propagation algorithm variations. Dilated convolutions are utilized to create hierarchical temporal models that are able to capture the long term temporal dependencies of input data through a large receptive field. The basic concept is to build a stack of dilated convolutional layers, with each layer adding to the stack's dilation rate. It is usual practice of starting the network's first layer at a initial dilation rate of $q = 1$, and then double the $q$ value for each subsequent layer. Despite pooling or strided convolutions, this method exponentially expands the model's receptive field without decreasing the resolution. For both causal and a causal situations, with $L>0$ number of layers, the value of the receptive field is given as follows:
\begin{equation}
    r_{causal} = 2^{(L-1)} \times k
\end{equation}
\begin{equation}
    r_{acausal} = 1 + \frac{k}{2}(2^{L+1} - 2)
\end{equation}

In conclusion, the convolution layers are characterized by three parameters: the number of filters $n_{filters}$, the dilation rate $q$, and the length of the filter (kernel size) $k$. Generally, a convolution layer maps a input sequence $\mathbf{x}: \mathbb{T} \rightarrow \mathbb{R}^d$ to the sequence of output $\mathbf{y}: \mathbb{T} \rightarrow \mathbb{R}^{n_{filters}}$, where the shape of output is independent of k.

\begin{figure}
    \centering
    \includegraphics[scale = 0.15]{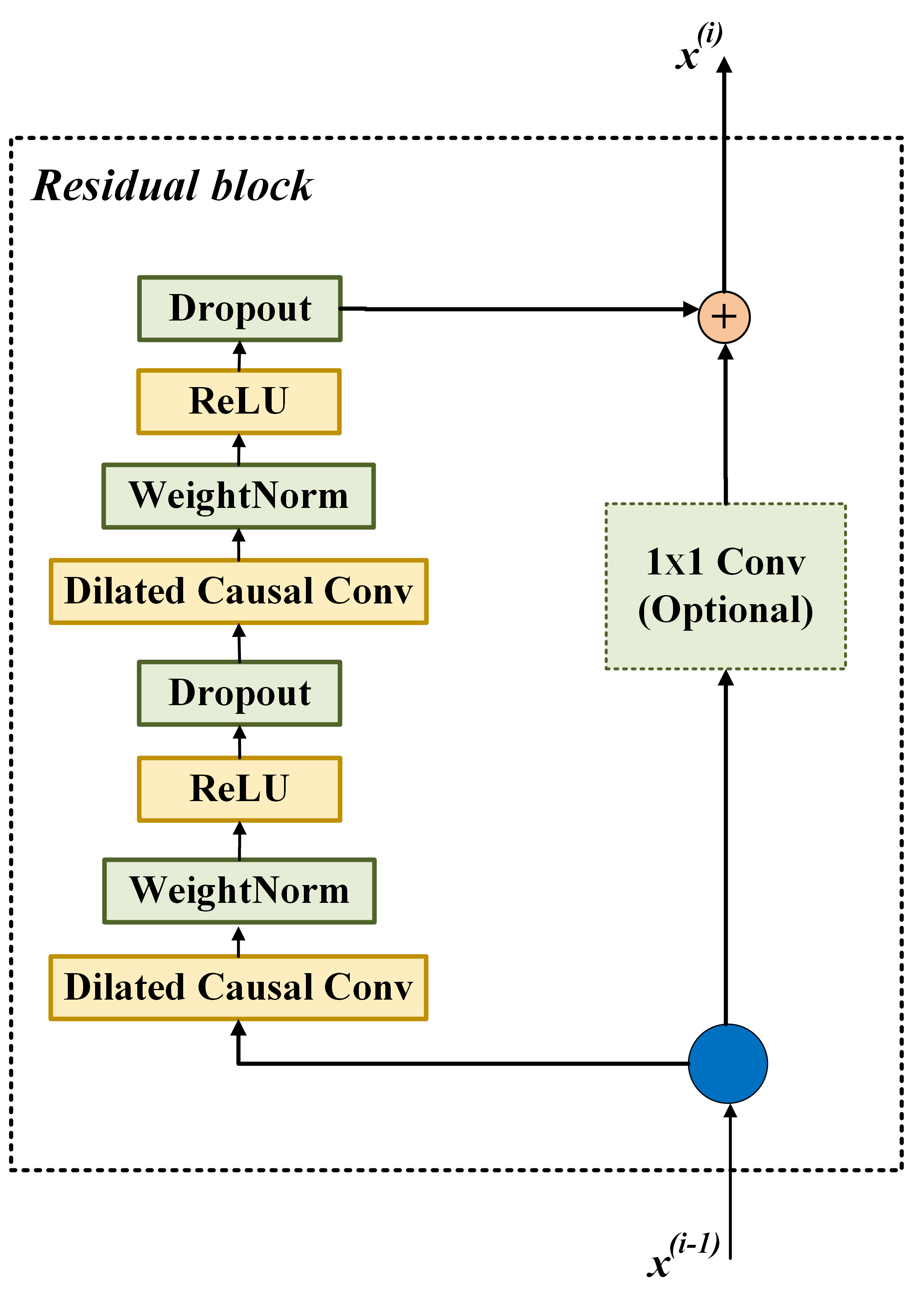}
    \caption{Residual block of TCN}
    \label{fig: residual block}
\end{figure}

\subsubsection{Temporal Convolutional Networks}

Although several conventional architectures inspire the TCN \cite{archive}, it differs from them as it integrates long term memory, auto regressive prediction, simplicity, and residual blocks. The TCN's description is illustrated as a successive chain of two sub-blocks that are combined sequentially from $n$ residual blocks. A sequence of dilated convolution layers, weighted normalization layers, rectified linear units (ReLU) activation functions, and spatial dropout layer sequences make up each subblock of TCN. To bypass the residual blocks, a skip connection is additionally added to its output. 
The three characteristic parameters typically used to describe the TCN are: list of dilation rates ($q_1, q_2,.....q_L$), kernel size $k$, and the number of filters $n_{filters}$. The sequence $\textbf{y}: \mathbb{T} \rightarrow \mathbb{R}^{n_{filters}}$ is the output for each residual block, and also for the final ouptut. A residual block of baseline TCN is depicted in Fig. \ref{fig: residual block}. Where, each block of TCN has two dilated causal convolutions layers, for which ReLU function \cite{nair2010rectified} is used as an activation functions. Weighted normalization \cite{salimans2016weight} is used on the convolution layers for normalizing them. In addition, each dilated convolution layer is followed by a spatial dropout \cite{srivastava2014dropout} for regularization.

\subsubsection{Autoencoder using TCNs}

\begin{figure}[h]
    \centering
    \includegraphics[scale = 0.12]{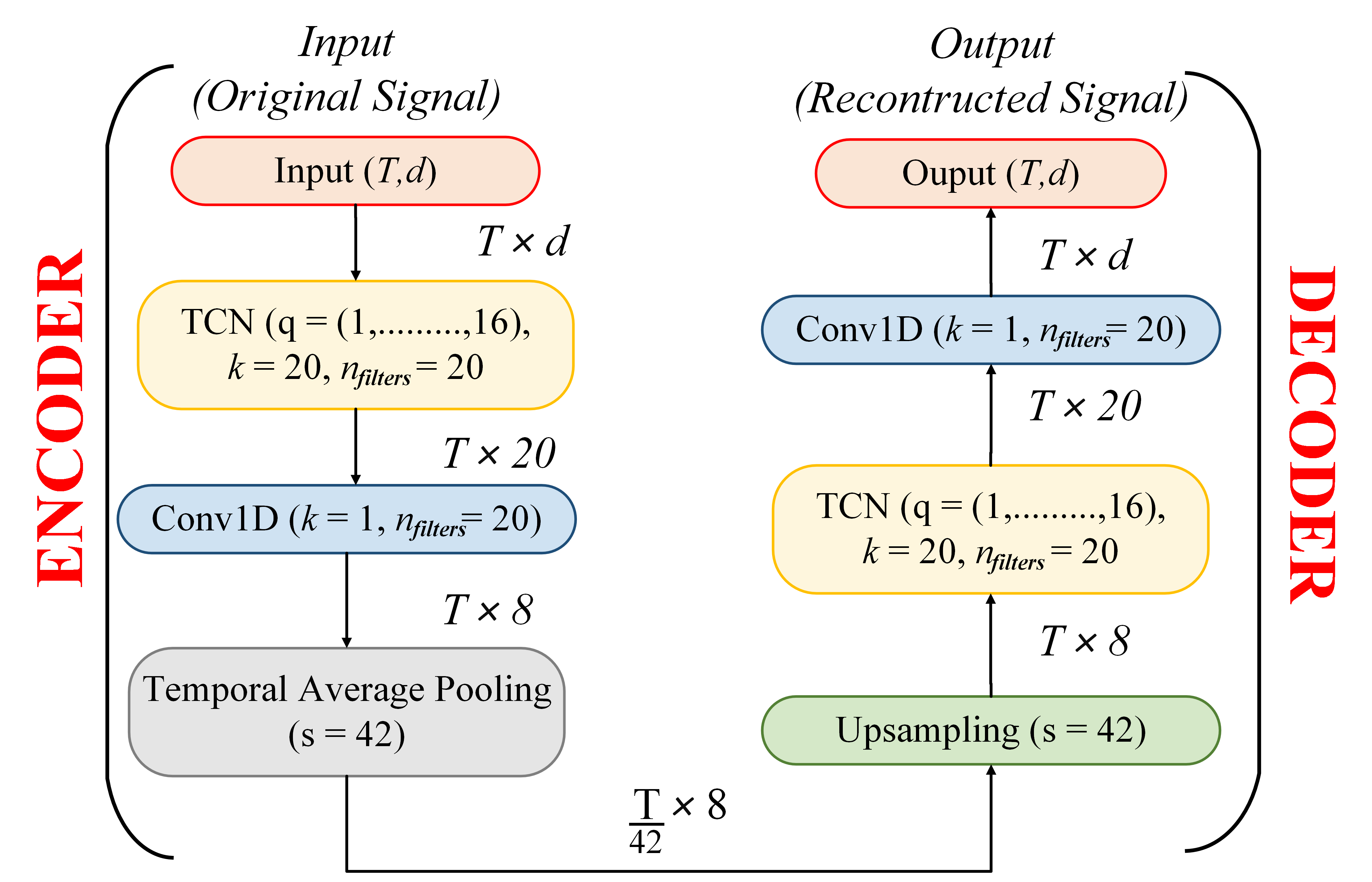}
    \caption{TCN-AE architecture describing the parameters of each layer inside the box. Sequence $x[n]$ with dimensionality $d$ and length $T$ is the input to the TCN-AE \cite{conference}.}
    \label{fig: architecture}
\end{figure}

The autoencoder using TCNs is named as TCN-AE, and its architecture is shown in Fig. \ref{fig: architecture}. Like other autoencoders, the TCN-AE consists of an encoder and decoder. Initially, the encoder $enc(.)$ processes the input signal $\textbf{x}[n]$ of dimension $d$ and length $T_{train}$ using TCN. It makes an effort to provide a compact representation that highlights the key elements of the input sequences and permits a passably accurate reconstruction in a subsequent stage. Finding both short and long term dependencies is essential for learning the important features of sequence. The encoder passes the input sequence through the TCN network to obtain the feature map. This feature map is reduced (dimension reduction) by applying the output to a one-dimensional $1 \times 1$ convolutional layer with $k=1$, $q=1$, and a small number of filters $c$. The final layer of the encoder, called temporal average pooling, is in the process of downscaling the series by a factor of $s$. This was achieved by averaging the series in groups of size $s$ along the temporal axis. Where the dimension of encoded representiation is specified by the number of filters ($n_{filters}$) and the sampling factor $s$ specifies the factor amount by which the time series $T$ has to be reduced. Thus, the encoded representation is generated by compressing the original input $\textbf{x}[n]$ and is represented as $\textbf{g}[n] = enc(\textbf{x}[n])$, where $\textbf{g}: \{ 0,1,....,T/s-1 \} \rightarrow \mathbb{R}^c$.

By utilizing encoder's output as an input, the decoder $dec(.)$ tries to reconstruct the original input sequence. This output from the encoder first passes through the upsampling layer, which duplicates each point in the series $s$ times by performing a simple sample-and-hold neighbor interpolation. The upscaled sequence is passed to a second TCN block structurally identical to the encoder's TCN block but with independent and united weights. Finally, the original dimension is restored through another $1 \times 1$ convolutional layer with d filters and is used to obtain the reconstructed output $\hat{\textbf{x}}[n] = dec(\textbf{g}[n])$, $\hat{\textbf{x}}: \mathbb{T} \rightarrow \mathbb{R}^d$ \cite{elsiever}.

\subsection{Anomaly detection using TCN-AE}

% TCN-AE provides a powerful and flexible approach to anomaly detection in time series by utilizing the TCN capabilities to capture the temporal dependencies and the autoencoders to learn the underlying data patterns. When the TCN-AE model is trained on time series data predominantly composed of nominal patterns, it strives to minimize the reconstruction error associated with these nominal sequences. Simultaneously, the model expects the reconstruction error to be larger for anomalous patterns or those exhibiting significant deviations in their characteristics. One of the possibilities for identifying the anomalous patterns is estimating the distribution of the reconstruction error. In this paper, 

The anomalies using TCN-AE are detected by using a sliding window approach, where a window of length $l$ is slid over reconstruction error and a covariance matrix $\Sigma_{cov}$ and a mean vector $\mu$ are computed. Subsequently, the Mahalanobis distance \cite{de2000mahalanobis} is used as an anomaly score. Algorithm 1 provides a uniform algorithmic explanation and the anomaly detection process using TCN-AE \cite{conference}.

\begin{algorithm*}
\caption{Algorithm for anomaly detection}
\label{alg:anomaly_detection}
\begin{algorithmic}[1]
\State \textbf{Tunable parameters:}
\State $M_T\text{: Anomaly detection threshold}$
\State $\textit{l}\text{ : length of the window for error vectors construction}$
\State $T_{\text{train}}$ : training sub-sequences length
\Function{detectanomaly}{$x_{\text{tr}}[n]$, $x[n]$} \Comment{$x_{\text{tr}}$, $x : \mathbb{N} \to \mathbb{R}^d$ of length $T$}
    \State \textbf{Function} $\text{TCNAE}()$ model \& \textbf{Initialize} training parameters
    \For{$\{1....n_{\text{epochs}}\}$}
        \State Extract sub-sequences $X_{\text{train}}^{(i)} \in \mathbb{R}^{T_{\text{train}} \times d}$ for training from $x_{\text{tr}}[n]$, where $i = 1, \ldots, B$
        \For{$i \in \{1, \ldots, B\}$}
            \State Train $\text{TCNAE}$($X_{\text{train}}^{(i)}$) \Comment{Network training on mini-batches}
        \EndFor
    \EndFor
    \State $\hat{x}[n] = \text{TCNAE}(x[n])$ \Comment{Encoding $\&$ reconstructing the whole sequence}
    \State \text{error}$[n] =$ \text{Difference}($x[n], \hat{x}[n]$) \Comment{reconstruction error of length $T$, $e : \mathbb{N} \to \mathbb{R}^d$}
    \State $E[n] = \text{reshapedslidingwindow}(\text{error}[n], \textit{l})$ \Comment{$E' : \mathbb{N} \to \mathbb{R}^{T \times '} \times d$}
    % \State $E'[n] = \text{reshape}(E[n])$ \Comment{$E_0 : \mathbb{N} \to \mathbb{R}^{T \times '} \times d$}
    \State $\Sigma_{cov}, \mu = \text{estimate}(E'[n])$ \Comment{$\mu \in \mathbb{R}^{'} \times d$, $\Sigma_{cov} \in \mathbb{R}^{'} \times d \times d$}
    \State $M[n] = (E'[n] - \mu)^T \Sigma_{cov}^{-1} (E'[n] - \mu)$ \Comment{Anomaly score calculation} %from Mahalanobis distance
    \State $a[n] = \begin{cases} 
                    1 & \text{if } M[n] \geq \textit{M}_T \\
                    0 & \text{else}
                  \end{cases}$ \Comment{Anomaly flags (Binary)}
    \State \Return $a[n]$ \Comment{For the entire time series return binary anomaly flags}
\EndFunction
\end{algorithmic}
\end{algorithm*}

The attacked data ($x[n]$) is fed to the trained model for obtaining the reconstructed data ($\hat{x}[n]$). Then the error ($e$) between both data is computed. By using a sliding window approach, the covariance matrix ($\Sigma_{cov}$) and the mean vector ($\mu$) are computed for each error window ($E[n]$) of length $l$.

Once the covariance matrix ($\Sigma_{cov}$) and the mean vector ($\mu$) are obtained, the anomaly score is computed with them and the error window. The anomaly score is obtained as the Mahalanobis distance ($M[n]$) and is computed as follows:
\begin{equation}
    M[n] = (E[n] - \mu)' \Sigma^{-1} (E[n] - \mu)
\end{equation}

The anomaly threshold ($M_T$) is obtained by considering a subsequence of 10$\%$ of the data, by optimizing it on this short subsequence. The threshold thus optimized is compared with the anomaly score ($M[n]$) of each time series window. The algorithm detects that an anomaly is present, if the anomaly score is greater than $M_T$, else the algorithm treats it as no anomaly.

% \subsection{Determining Anomaly detection threshold} 

The output for each point of the time series data is the anomaly score. A higher anomaly score signifies that anomalies are present while the lower anomaly score indicates that anomalies are absent. An anomaly threshold is required to compare the anomaly score for classifying each point as an abnormal or normal point. Points with scores above the threshold are called anomalous and those with low scores are called normal points. The threshold is determined by analyzing a sub-sequence comprising 10$\%$ of the dataset. The Anomaly threshold is optimized within this subset with the aim of maximizing the F1 score. Subsequently, the optimal threshold is extrapolated to the entire time series to derive the final outcomes.

% \section{Preliminaries}
% \label{Section: Preliminaries}
\section{Simulation and Methodology}
\label{Section: System Modelling and Data Acquisition}

\subsection{System Modelling}

\begin{figure}[h]
    \centering
    \includegraphics[scale=0.13]{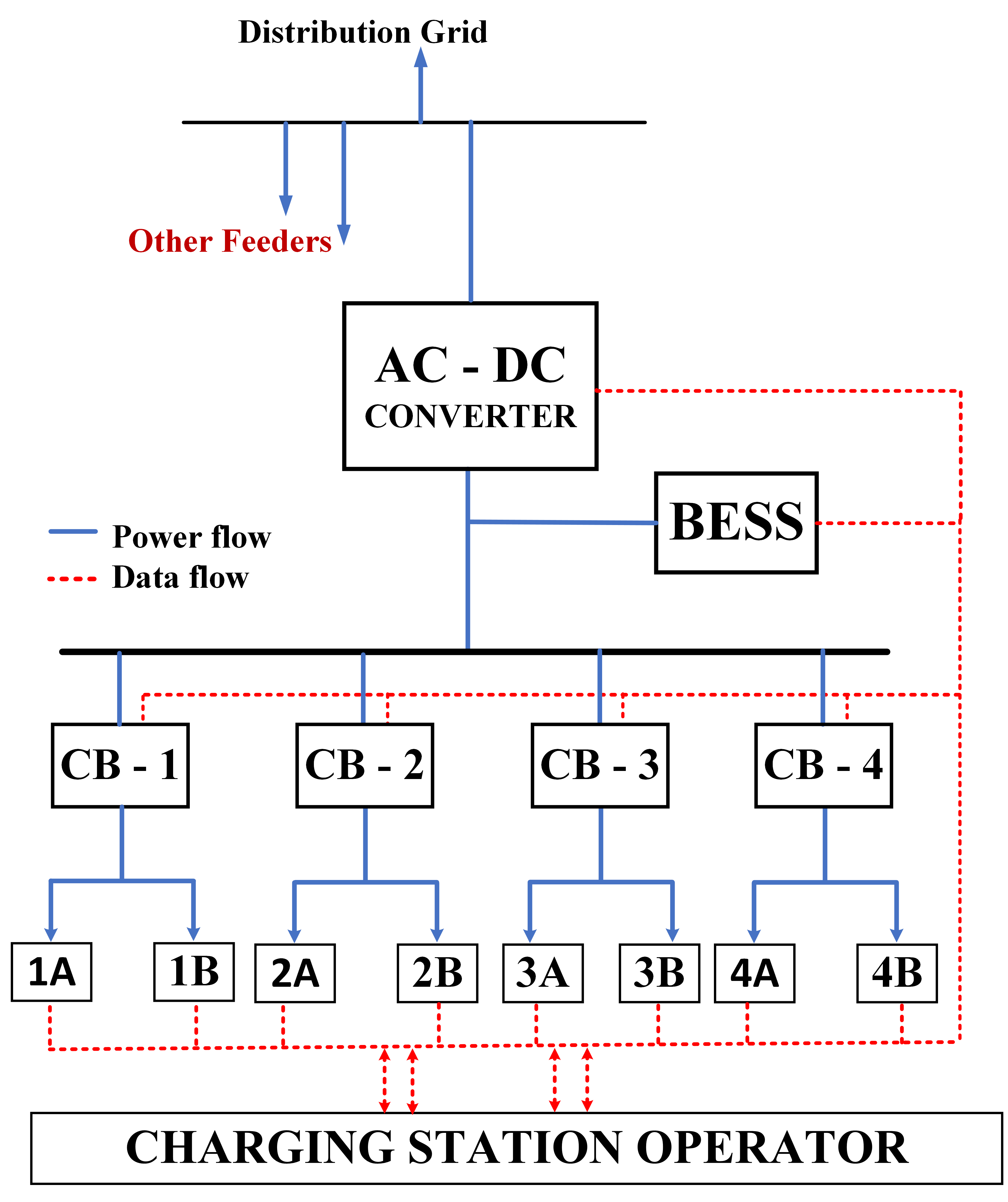}
    \caption{Proposed architecture of EVCI}
    \label{fig: EVCI}
\end{figure}

The considered EVCI model consists of four charging boards (CB$_{i}$, where $i = 1,2,3,4$) with each board supporting two charging ports denoted by $CB_{\alpha \beta}$ with $\alpha$ represents the CB number and $\beta$ desingates the port identifier ($A$ or $B$). To facilitate load balancing, a BESS is integrated into the system, which is obtained by charging and discharging it during the normal and power shortage durations respectively. Fig. \ref{fig: EVCI} illustrates the overall architecture of the EVCI system.

% The proposed EVCI consists of four charging boards (CB$_{i}$, $i = 1,2,3,4$), where two charging ports are connected to each CB. The charging ports are represented as CB$_{iy}$, where $i$ indicates the CB number and $y$ is the charging port representation, i.e., either A or B. Further, a Battery Energy Storage System (BESS) is also connected to the system for load balancing. The function of the BESS is to supply the power for charging the EVs whenever there is a deficit from the grid and it gets charged during normal conditions. The detailed architecture of the EVCI system is shown in Fig. \ref{fig: EVCI}. 

The charging systems are developed in compliance with IEC and BIS \cite{chargestd} to accommodate both four-wheeler and two-wheeler vehicles at various charging rates. Table \ref{tab: EVCI configuration} shows the type of charging port a corresponding EV can be charged. The system incorporates two categories of DC fast chargers that are classified as Level-1 and Level-3, whose ratings are 50kW, 100V and 400kW and 400V respectively. 

% The EVCI system is designed for charging the both 2 wheeler and 4 wheeler vehicles at various charging rates. The type of the vehicle that can be charged at the respective ports are mentioned in Table \ref{tab: EVCI configuration}. Two types of power level DC chargers are designed in the system namely level-1 and level-3. The level-1 DC chargers are rated for 50 kW and 100 volts, whereas the level-2 DC chargers are rated for 400 kW and 400 volts. These chargers are designed according to the BIS and International Electrotechnical Commission (IEC) Standards \cite{chargestd}. 

The EV battery specification considered for two-wheeler EVs is 72V and 45Ah to access charging ports $CB_{1A}$ and $CB_{1B}$, and for four-wheelers, two specifications are considered: 200V, 90Ah for Level-1 and 300V, 100Ah for Level-3 charging ports respectively. Multiple operating scenarios are considered and simulated, such as i) BESS is charging and the grid is solely supplying power, ii) the grid is unable to power the grid and the BESS is used to charge the EVs, iii)Both the grid and EVs are used to charge the EVs. As this study does not encompass advanced smart grid functionalities, such as demand response, the analysis excludes Vehicle-to-Grid (V2G) operations. Instead, the focus is solely on Grid-to-Vehicle (G2V) energy transfer \cite{wu2022hierarchical}. 

% In the considered architecture of the EVCI, a specific category of EVs comprising 2-wheelers with battery specifications up to 72 volts and 45 Ah are designated to access charging ports CB$_{1A}$ and CB$_{1B}$. Meanwhile, two distinct categories of 4-wheelers are considered: those with battery specifications of up to 200 volts and 90 Ah are directed to utilize level-1 charging ports, while vehicles with battery specifications of up to 300 volts and 100 Ah are directed to utilize level-3 charging ports. The EVCI is simulated for various cases such as i) only the grid supplying power while the BESS is getting charged, ii) the BESS is charging the EVs while the grid is unable to supply the power, and iii) both the grid and BESS is used for charging the vehicles. Since the modern operations of the smart grids, like demand response, are out of the scope of this work, the Vehicle-to-Grid (V2G) operations are not considered and only the Grid-to-Vehicle (G2V) operations are considered \cite{wu2022hierarchical}.

It is assumed that the Charging Station Operator (CSO) receives each component's parametric data of the EVCI, whose responsibilities are infrastructure management and development, reporting, data analysis, payments, and regulatory compliance, etc., \cite{chen2020review}. However, the primary aim of this work is anomaly detection of charging port current data; only the electrical parameters such as voltage, current, and charge status of each charging port are considered along with the grid and battery current. Furthermore, the charging port upon connecting the EV shares the information such as the arrival SoC of EV, desired level of SoC at the time of leaving, departure time, etc.The following sections elaborate on how these data parameters are processed and made appropriate for anomaly identification utilizing the suggested methodology.

% The proposed architecture presupposes that the Charging Station Operator (CSO) receives comprehensive data encompassing various parameters from each component. The CSO undertakes various responsibilities, including infrastructure management and development, data analysis and reporting, emergency and demand response, payment processes, regulatory compliance, etc., \cite{chen2020review}. Since detecting anomalies in the charging port current values is the primary objective of this work, it is assumed that the CSO collects data comprising parameters such as current, voltage, and charge status from each CB and charging port. Additionally, data regarding the magnitude of current drawn from both the battery and grid sources is gathered to discern the power supply source. Furthermore, upon the arrival of an EV, communication between the EV and CSO includes details such as SoC as well as departure and arrival time requirements, indicating the desired departure time and the desired SoC level for charging the EV. 

\begin{table}[h]
    \centering
    \caption{EVCS Configuration}
    \renewcommand*{\arraystretch}{1.5}
    \label{tab: EVCI configuration}
    \begin{tabular}{>{\centering\arraybackslash}p{1.5cm}>{\centering\arraybackslash}p{1.5cm}>{\centering\arraybackslash}p{2.5cm}}
    \toprule
    \textbf{Charging Board} & \textbf{Charging Port} & \textbf{Vehicle and Charger type} \\
    \midrule
    \multirow{2}{*}{CB$_{1}$} & CB$_{1A}$ & Level-1, 2-Wheeler\\    
                               & CB$_{1B}$ & Level-1, 2-Wheeler\\
    \midrule
    \multirow{2}{*}{CB$_{2}$} & CB$_{2A}$ & Level-1, 4-Wheeler \\
                               & CB$_{2B}$ & Level-1, 4-Wheeler\\
    \midrule
    \multirow{2}{*}{CB$_{3}$} & CB$_{3A}$ & Level-1, 4-Wheeler \\
    
                               & CB$_{3B}$ & Level-3, 4-Wheeler\\
    \midrule
    \multirow{2}{*}{CB$_{4}$} & CB$_{4A}$ & Level-3, 4-Wheeler\\
    
                               & CB$_{4B}$ & Level-3, 4-Wheeler \\
    \bottomrule
    \end{tabular}
\end{table}

\subsection{Methodology}
\label{Section: Methodology}
The workflow of methodology for identifying the anomalies in the charging data of EVCI is shown in Fig. \ref{fig: workflow}. The anomalous data is created by replacing the charging current magnitudes, of the non-charging periods with the charging periods and vice-versa. A binary variable named 'Anomaly' is created where the variable's value is $1$ for these anomalous points and $0$ elsewhere. The data thus obtained is fed to the TCN-AE model for training purposes and reconstructed data is obtained. The anomaly score and anomaly signal are then calculated using a sliding window technique. Furthermore, the detailed explanation of the data preprocessing and the anomaly detection procedure are explained below.

% The proposed methodology for creating and detecting the anomalies is shown in Fig. \ref{fig: workflow}. The LSTM based autoencoder is trained with preprocessed data, and FGSM is implemented on the same model by computing the gradients for loss functions of specific sequential input datasets. With the help of the gradients, the perturbations are added to these input datasets and then replaced with the original datasets with their corresponding time indices. These new input datasets are fed to the LSTM autoencoder to predict the output. This output may contain the anomalies for the instances of the perturbed inputs. A sliding window approach is then used in comparing the distributions of predicted values and observed values for a particular window length and then compared with the threshold to detect anomalies in the charging port current magnitudes.

\subsubsection{Data Preprocessing}

The data acquired by simulating the proposed EVCI on various scenarios are in raw format and must be preprocessed to make them suitable for the neural network model for efficient learning and prediction. This data contains the parameters related to the eight charging ports of the system. The model is used to predict the current magnitude of charging port A ($I_{CB_{1A}}$), by feeding voltages of charging ports ($V_{CB_{1A}}$. $V_{CB_{2B}}$) associated with the $CB_1$, EVS CS value ($CS_{EV_{1A}}$, $CS_{EV_{1B}}$), and the current magnitudes of grid and BESS ($I_{grid}$, $I_{BAT}$). The effect of different scaled values on the model performance, such as biasing, and vanishing gradients, is eliminated by scaling all the input values to a common scale. This is used by using a standard scaler \cite{zhu2016influence} method. The scaled datasets are further partitioned into training and testing datasets for model training and validation.

\subsubsection{Proposed Model and Methodology}

\begin{figure}[]
    \centering
    \includegraphics[scale = 0.2]{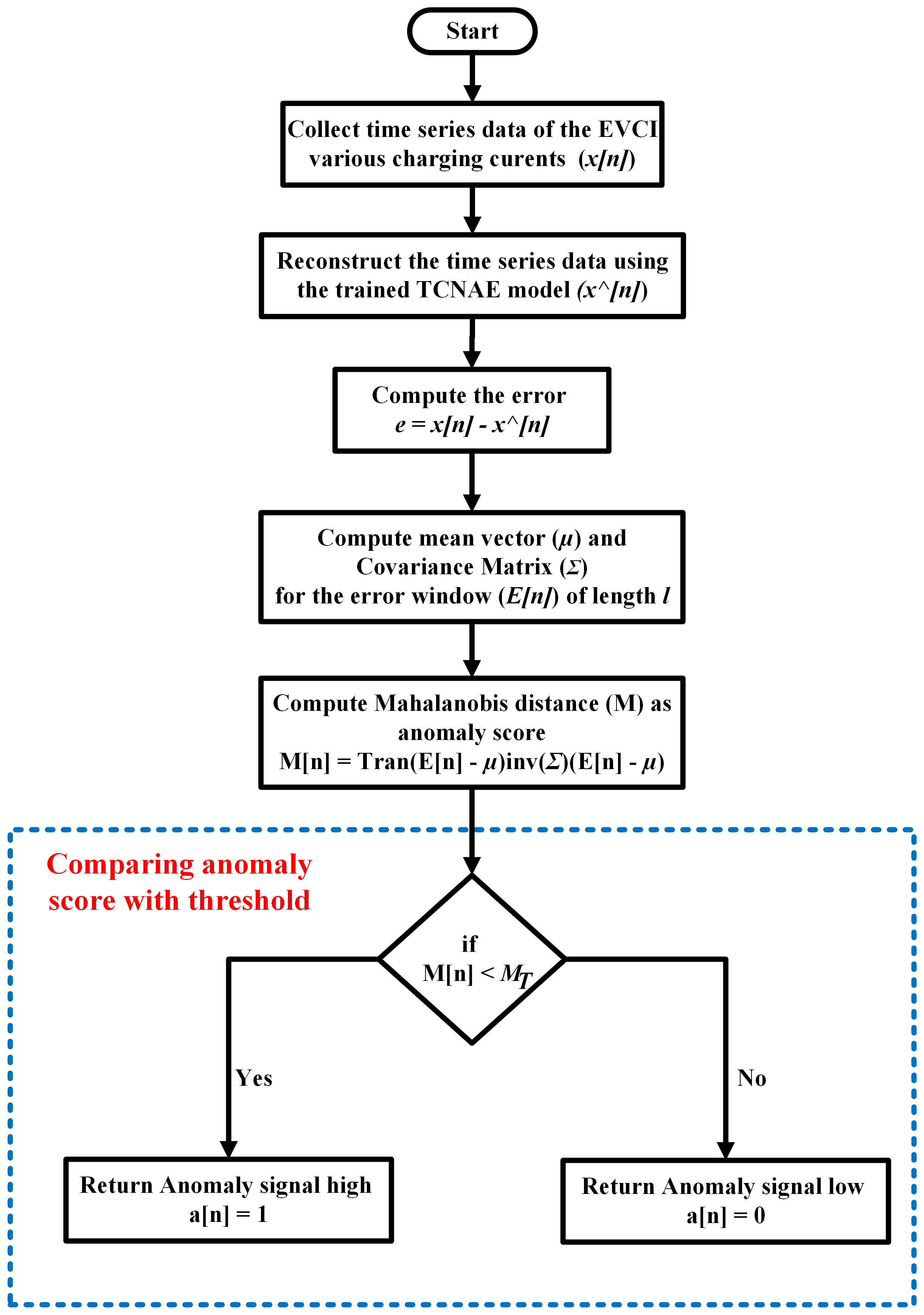}
    \caption{Flowchart for proposed Anomaly Detection}
    \label{fig: flowchart}
\end{figure}

The proposed work uses a TCN-AE model for training the model with the sequential multi variate time series data. The sequential architecture of the model i sshown in Fig. \ref{fig: architecture}. The architecture of the model encompassing each layer, with number of nodes, filters etc., are provided in Table \ref{tab: Model Parameters}. The parameters are fine-tuned using GridSearchCV to identify the optimal model configuration. The trained model is subsequently provided with data containing anomalies for reconstruction. Fig. \ref{fig: flowchart} depicts the detailed procedure of the proposed anomaly detection used in this work.

% The proposed work uses an LSTM-based autoencoder cascaded with hidden layers, for forecasting the charging port current magnitudes. Fig. \ref{fig: NN architecture} shows the architecture of the model with the layers sequentially, along with the dropout and repetition-vector layers. The architectural specifications of the neural network, encompassing the number of hidden layers and their respective units, the activation and loss functions employed, the optimizer utilized for model convergence, and the learning rate value during training, are outlined in Table \ref{tab: parameters}. These parameters are optimized via GridSearchCV to determine the most effective configuration for the proposed model. This trained model is then fed with the adversarial inputs generated by the FGSM as shown in Fig. \ref{fig: workflow}. 

\begin{table}[h]
    \centering
    \caption{Parameters of the TCN-AE Model}
    \label{tab: Model Parameters}    
    \begin{tabular}{>{\centering\arraybackslash}p{3.0cm}>{\centering\arraybackslash}p{2.0cm}}
    \toprule
    \textbf{Parameter} & \textbf{Value}\\
    \midrule
    Optimizer & Adam\\
    Learning rate & 0.001\\
    Loss & logcosh\\
    Activation Function & Linear\\
    Number of filters ($n_{filters}$) & 20\\
    Dilation rate ($q$) & (1,2,....,16)\\
    \bottomrule
    \end{tabular}
\end{table}

\section{Results and Discussions}
\label{Section: Results and Discussions}

\begin{figure}[h]
    \centering
    \includegraphics[scale=0.4]{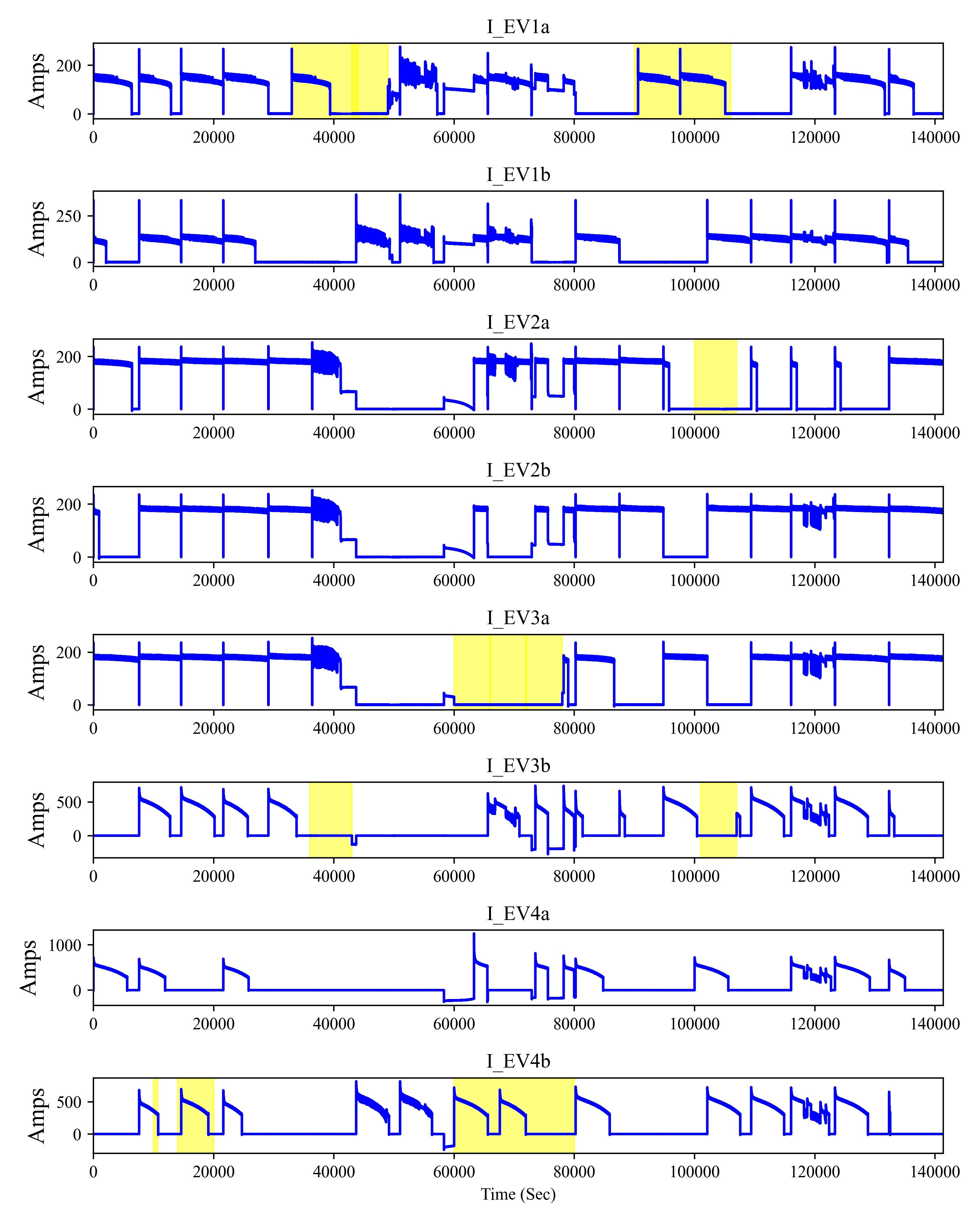}
    \caption{Various charging port current magnitudes highlighting the spoofed data points}
    \label{fig: dataa}
\end{figure}

The proposed EV charging station is developed in MATLAB/Simulink environment and analyzed under various operational scenarios. The simulation considers different power sources, including grid power alone, battery energy storage system (BESS) power alone, and a combination of both. Additionally, multiple EVs are charged at distinct ports to generate training and testing datasets for further analysis. The charging port current magnitude values, highlighting the spoofed locations are represented in Fig. \ref{fig: dataa}. This provides the visualized representation of the multi-variable spoofing at the same time.

\begin{figure}[h]
    \centering
    \includegraphics[scale=0.5]{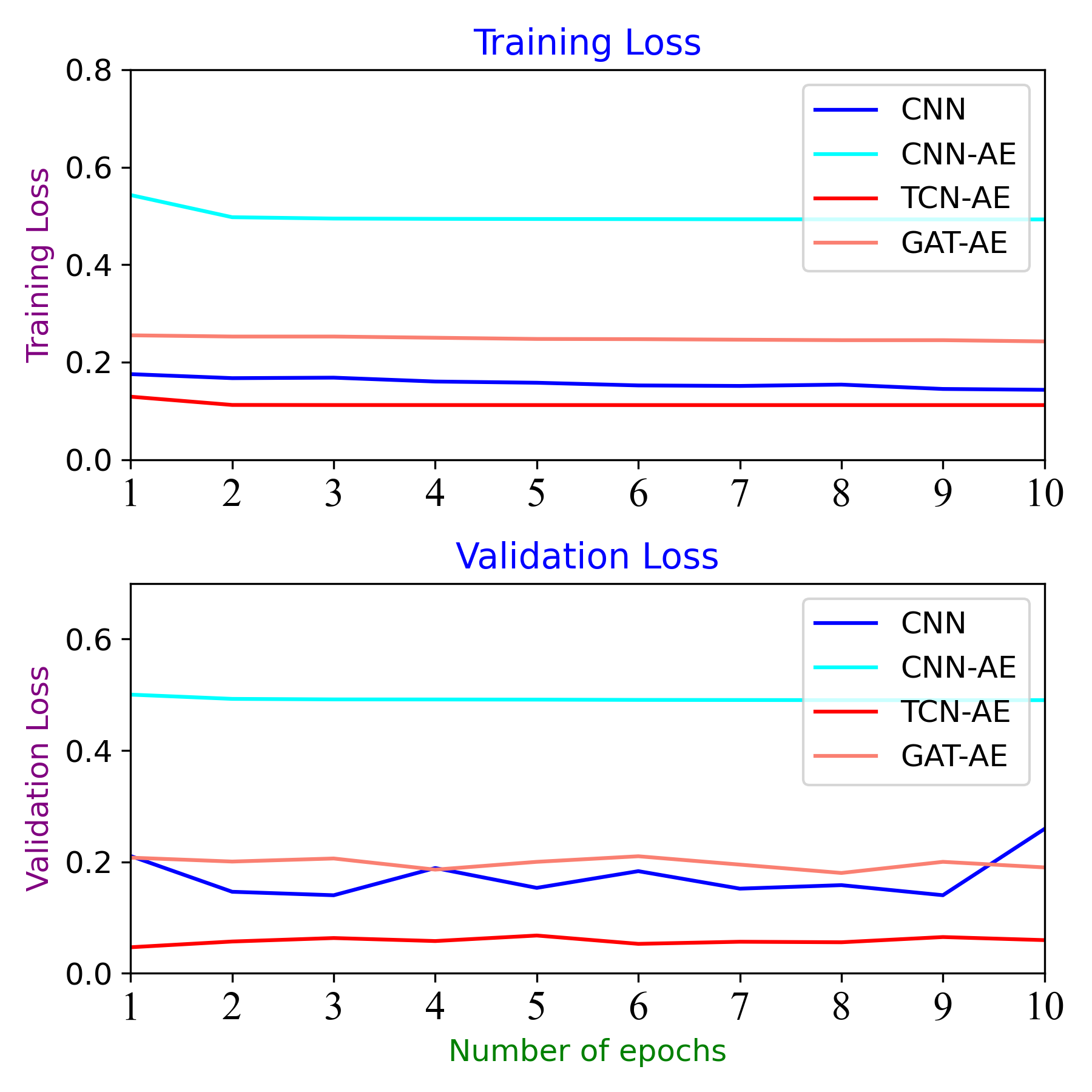}
    \caption{Performance of the various models over the proposed model}
    \label{fig: model evaluation}
\end{figure}

The TCN-AE model exhibits superior performance on the acquired dataset over the other models. Fig. \ref{fig: model evaluation} illustrates the model's effectiveness in reconstructing the original data and compares the loss function for 10 epochs and batch size 32 with the other models like CNN \cite{kwon2018empirical}, CNN-AE \cite{choi2022multivariate}, and Graph Attention Network based autoencoder (GAT-AE) \cite{zhao2020multivariate}.

\begin{table}[h]
    \centering
    \caption{Performance evaluation of the proposed methodology}
    \label{tab:metrics}    
    % \begin{tabular}{|c|c|c|c|c|}
    \begin{tabular}{>{\centering\arraybackslash}p{2.0cm}>{\centering\arraybackslash}p{1cm}>{\centering\arraybackslash}p{1cm}>{\centering\arraybackslash}p{1cm}>{\centering\arraybackslash}p{1cm}}
         \toprule
         Method & Accuracy & Precision & Recall & F1-score \\
         \midrule
         TCN-AE (proposed) & 0.9964 & 0.99521 & 0.9954 & 0.9944\\
         \midrule
         GAT-AE & 0.9614 & 0.95 & 0.94 & 0.96\\
         \midrule
         CNN  & 0.9894 & 0.9842 & 0.9836 & 0.983\\
         \midrule
         CNN-AE  & 0.956 & 0.9598 & 0.9553 & 0.956\\
         \bottomrule
    \end{tabular}
\end{table}

% \begin{table}[h]
%     \centering
%     \caption{Performance evaluation of the proposed methodology}
%     \label{tab:metrics}    
%     % \begin{tabular}{|c|c|c|c|c|}
%     \begin{tabular}{>{\centering\arraybackslash}p{2.0cm}>{\centering\arraybackslash}p{1cm}>{\centering\arraybackslash}p{1cm}>{\centering\arraybackslash}p{1cm}>{\centering\arraybackslash}p{1cm}}
%          \toprule
%          Method $\&$ Ref. & Accuracy & Precision & Recall & F1-score \\
%          \midrule
%          TCN-AE (proposed) & 0.9964 & 0.99521 & 0.9954 & 0.9944\\
%          \midrule
%          GAN-RF \cite{sun2022data} & 0.9614 & 0.95 & 0.94 & 0.96\\
%          \midrule
%          SVM \cite{hong2021svm} & 0.9894 & 0.9842 & 0.9836 & 0.983\\
%          \midrule
%          RF \cite{dixit2022anomaly} & 0.956 & 0.9598 & 0.9553 & 0.956\\
%          \bottomrule
%     \end{tabular}
% \end{table}
\subsection{Performance Evaluation}

The performance of the proposed anomaly detection method is quantified using the four metrics, namely; Accuracy, Precision, Recall, and F1-score, which are given as follows:
\begin{enumerate}
    \item \textbf{Accuracy} - It measures the overall correct of the anomaly detection method by comparing the number of correctly classifying instances with the total instances. A better model has a high accuracy value.
    \begin{equation}
    \text{Accuracy} = \frac{\mathcal{TP} + \mathcal{TN}}{\mathcal{TP} + \mathcal{TN} + \mathcal{FP} + \text{FN}}
    \end{equation}
    \item \textbf{Precision} - It measures the correctly classified instances to the overall anomaly classified instances. It helps in repeatability and reproducability of the model.
    \begin{equation}
        \text{Precision} = \frac{\mathcal{TP}}{\mathcal{TP} + \mathcal{FP}}
    \end{equation}
    \item \textbf{Sensitivity} - It measures the portion of correctly detected anomalies out of the total anomalies present in the data. It is also referred to as True Positive Rate (TPR). Higher recall rate indicates the effective method in identifying the majority of the anomalies, and reducing the false identification.   
    \begin{equation}
        \text{Recall} = \frac{\mathcal{TP}}{\mathcal{TP} + \mathcal{FN}}
    \end{equation}
    \item \textbf{F1-Score} - It represents the balance between precision and recall, aiding in simultaneously achieving high values for both metrics. It is computed as the harmonic mean between the precision and recall.
    \begin{equation}
        \text{F1-Score} = \frac{2 \times \text{Precision} \times \text{Recall}}{\text{Precision} + \text{Recall}}
    \end{equation}
\end{enumerate}

The parameters TP (True Positive), FP (False Positive), TN (true Negative), and FN (false Negative) are defined as follows:
\begin{itemize}
    \item $\mathcal{TP}$ - Instances correctly identified as anomalies.
    \item $\mathcal{FP}$ - Instances of benign data, incorrectly classified as anomalies.
    \item $\mathcal{TN}$ - Instances of benign data correctly classified as anomalies.
    \item $\mathcal{FN}$ - Instances of abnormal data incorrectly classified as normal.
\end{itemize}

Once the model is trained, the original dataset is fed to the model to obtain the reconstructed data. The reconstructed data is tested for proposed anomaly detection. The training data after feeding the model is fed back to the proposed anomaly detection and the anomaly signal detection is detected. The sliding window's length is specified as 128. The performance of the proposed model is evaluated and comparison of the metrics with the few state-of-art methods are shown in Table \ref{tab:metrics}. The results shows that the TCN-AE is superior to the other methods in terms of all the four metrics.

\section{Conclusions}
\label{Section: Conclusion}

A simulation model of an EVCI is developed to accommodate both two-wheeler and four-wheeler EVs by incorporating four CBs and two charging ports for each CB. The system is evaluated under multiple operational scenarios, with data collected and partitioned for training and testing. A TCN-AE model is employed to reconstruct the multivariate time-series data corresponding to charging port currents. To simulate replay attacks, anomalous data is introduced by substituting the actual charging port current values with data from different time instances of the same port.

The reconstructed data obtained from the TCN model is used for detecting the anomalies by computing the error between the original and reconstructed data. A sliding window approach is used to detect the anomalies by computing the Mahalanobis distance as an anomaly score. The possible future works include assessing the risks that arise due to charging the EVs with inappropriate charging rates, causing potential damage to EV batteries; studying the grid instabilities due to inaccurate current values causing instability in demand and supply; and calculating the financial loss caused due to fraudulent charging transactions or inaccurate billings resulting from spoofed current values, which may also be treated as energy theft. Since this article provides in the detection of anomalies in the whole multivariate time series data, the detection of an anomaly in the individual signal and reducing the effects caused due to these anomalies is treated as future work.

\bibliographystyle{IEEEtran}
\bibliography{main}

% Generated by IEEEtran.bst, version: 1.14 (2015/08/26)
\begin{thebibliography}{10}
\providecommand{\url}[1]{#1}
\csname url@samestyle\endcsname
\providecommand{\newblock}{\relax}
\providecommand{\bibinfo}[2]{#2}
\providecommand{\BIBentrySTDinterwordspacing}{\spaceskip=0pt\relax}
\providecommand{\BIBentryALTinterwordstretchfactor}{4}
\providecommand{\BIBentryALTinterwordspacing}{\spaceskip=\fontdimen2\font plus
\BIBentryALTinterwordstretchfactor\fontdimen3\font minus \fontdimen4\font\relax}
\providecommand{\BIBforeignlanguage}[2]{{%
\expandafter\ifx\csname l@#1\endcsname\relax
\typeout{** WARNING: IEEEtran.bst: No hyphenation pattern has been}%
\typeout{** loaded for the language `#1'. Using the pattern for}%
\typeout{** the default language instead.}%
\else
\language=\csname l@#1\endcsname
\fi
#2}}
\providecommand{\BIBdecl}{\relax}
\BIBdecl

\bibitem{solaymani2019co2}
S.~Solaymani, ``Co2 emissions patterns in 7 top carbon emitter economies: The case of transport sector,'' \emph{Energy}, vol. 168, pp. 989--1001, 2019.

\bibitem{CO2emission}
``Iea energy system and transport, 2024.'' \url{https://www.iea.org/energy-system/transport}.

\bibitem{EVsales}
``Ev volumes-global ev sales for 2023.'' \url{https://www.ev-volumes.com/}.

\bibitem{chakraborty2024planning}
Victor, P.~Chakraborty, M.~Pal \emph{et~al.}, ``Planning of fast charging infrastructure for electric vehicles in a distribution system and prediction of dynamic price,'' \emph{International Journal of Electrical Power \& Energy Systems}, vol. 155, p. 109502, 2024.

\bibitem{das2020electric}
H.~S. Das, M.~M. Rahman, S.~Li, and C.~Tan, ``Electric vehicles standards, charging infrastructure, and impact on grid integration: A technological review,'' \emph{Renewable and Sustainable Energy Reviews}, vol. 120, p. 109618, 2020.

\bibitem{hamdare2023cybersecurity}
S.~Hamdare, O.~Kaiwartya, M.~Aljaidi, M.~Jugran, Y.~Cao, S.~Kumar, M.~Mahmud, D.~Brown, and J.~Lloret, ``Cybersecurity risk analysis of electric vehicles charging stations,'' \emph{Sensors}, vol.~23, no.~15, p. 6716, 2023.

\bibitem{industrydigits}
``Industry digits, automation strategies.'' \url{https://industrydigits.com/cyber-attacks-ev-charging-stations/}.

\bibitem{acharya2020cybersecurity}
S.~Acharya, Y.~Dvorkin, H.~Pand{\v{z}}i{\'c}, and R.~Karri, ``Cybersecurity of smart electric vehicle charging: A power grid perspective,'' \emph{IEEE Access}, vol.~8, pp. 214\,434--214\,453, 2020.

\bibitem{nasr2022power}
T.~Nasr, S.~Torabi, E.~Bou-Harb, C.~Fachkha, and C.~Assi, ``Power jacking your station: In-depth security analysis of electric vehicle charging station management systems,'' \emph{Computers \& Security}, vol. 112, p. 102511, 2022.

\bibitem{ahmed2016electric}
S.~Ahmed and F.~Dow, ``Electric vehicle and charging station technology as vulnerabilities threaten and hackers crash the smart grid,'' \emph{International Journal of Innovative Science, Engineering \& Technology}, 2016.

\bibitem{elsaeidy2021hybrid}
A.~A. Elsaeidy, A.~Jamalipour, and K.~S. Munasinghe, ``A hybrid deep learning approach for replay and ddos attack detection in a smart city,'' \emph{IEEE Access}, vol.~9, pp. 154\,864--154\,875, 2021.

\bibitem{ramanan2021blockchain}
P.~Ramanan, D.~Li, and N.~Gebraeel, ``Blockchain-based decentralized replay attack detection for large-scale power systems,'' \emph{IEEE Transactions on Systems, Man, and Cybernetics: Systems}, vol.~52, no.~8, pp. 4727--4739, 2021.

\bibitem{taher2023analyzing}
M.~A. Taher, M.~Tariq, M.~Behnamfar, and A.~I. Sarwat, ``Analyzing replay attack impact in dc microgrid consensus control: Detection and mitigation by kalman-filter-based observer,'' \emph{IEEE Access}, 2023.

\bibitem{le2024comprehensive}
Q.~Le~Roux, E.~Bourbao, Y.~Teglia, and K.~Kallas, ``A comprehensive survey on backdoor attacks and their defenses in face recognition systems,'' \emph{IEEE Access}, 2024.

\bibitem{humayed2017cyber}
A.~Humayed, J.~Lin, F.~Li, and B.~Luo, ``Cyber-physical systems security—a survey,'' \emph{IEEE Internet of Things Journal}, vol.~4, no.~6, pp. 1802--1831, 2017.

\bibitem{van2019non}
P.~Van~Aubel, E.~Poll, and J.~Rijneveld, ``Non-repudiation and end-to-end security for electric-vehicle charging,'' in \emph{2019 IEEE PES Innovative Smart Grid Technologies Europe (ISGT-Europe)}.\hskip 1em plus 0.5em minus 0.4em\relax IEEE, 2019, pp. 1--5.

\bibitem{chandwani2020cybersecurity}
A.~Chandwani, S.~Dey, and A.~Mallik, ``Cybersecurity of onboard charging systems for electric vehicles—review, challenges and countermeasures,'' \emph{IEEE Access}, vol.~8, pp. 226\,982--226\,998, 2020.

\bibitem{mitikiri2023modelling}
S.~B. Mitikiri, K.~Babu, D.~Dwivedi, V.~L. Srinivas, P.~Chakraborty, P.~K. Yemula, and M.~Pal, ``Modelling of the electric vehicle charging infrastructure as cyber physical power systems: A review on components, standards, vulnerabilities and attacks,'' \emph{arXiv preprint arXiv:2311.08656}, 2023.

\bibitem{farhadi2022charging}
P.~Farhadi and S.~M. Moghaddas~Tafreshi, ``Charging stations for electric vehicles; a comprehensive review on planning, operation, configurations, codes and standards, challenges and future research directions,'' \emph{Smart Science}, vol.~10, no.~3, pp. 213--245, 2022.

\bibitem{mavikumbure2023physical}
H.~S. Mavikumbure, V.~Cobilean, C.~S. Wickramasinghe, T.~Phillips, B.~J. Varghese, B.~Carlson, C.~Rieger, T.~Pennington, and M.~Manic, ``Physical anomaly detection in ev charging stations: Physics-based vs resnet ae,'' in \emph{2023 IEEE 32nd International Symposium on Industrial Electronics (ISIE)}.\hskip 1em plus 0.5em minus 0.4em\relax IEEE, 2023, pp. 1--7.

\bibitem{girdhar2021hidden}
M.~Girdhar, J.~Hong, H.~Lee, and T.-J. Song, ``Hidden markov models-based anomaly correlations for the cyber-physical security of ev charging stations,'' \emph{IEEE Transactions on Smart Grid}, vol.~13, no.~5, pp. 3903--3914, 2021.

\bibitem{li2020detecting}
Y.~Li, L.~Zhang, Z.~Lv, and W.~Wang, ``Detecting anomalies in intelligent vehicle charging and station power supply systems with multi-head attention models,'' \emph{IEEE Transactions on Intelligent Transportation Systems}, vol.~22, no.~1, pp. 555--564, 2020.

\bibitem{mitikiri2025anomaly}
S.~B. Mitikiri, V.~L. Srinivas, and M.~Pal, ``Anomaly detection of adversarial cyber attacks on electric vehicle charging stations,'' \emph{e-Prime-Advances in Electrical Engineering, Electronics and Energy}, vol.~11, p. 100911, 2025.

\bibitem{6023193}
H.~Liu, J.~Wu, and L.~Chu, ``Design of remote monitoring and fault diagnosis system for electric vehicle,'' in \emph{Proceedings of 2011 International Conference on Electronic \& Mechanical Engineering and Information Technology}, vol.~2, 2011, pp. 1079--1082.

\bibitem{yang2022electric}
Y.~Yang and J.~Li, ``Electric vehicle charging anomaly detection method based on multivariate gaussian distribution model,'' in \emph{Proceedings of the Asia Conference on Electrical, Power and Computer Engineering}, 2022, pp. 1--6.

\bibitem{basnet2020deep}
M.~Basnet and M.~H. Ali, ``Deep learning-based intrusion detection system for electric vehicle charging station,'' in \emph{2020 2nd International Conference on Smart Power \& Internet Energy Systems (SPIES)}.\hskip 1em plus 0.5em minus 0.4em\relax IEEE, 2020, pp. 408--413.

\bibitem{guo2020cyberattack}
L.~Guo, J.~Ye, and B.~Yang, ``Cyberattack detection for electric vehicles using physics-guided machine learning,'' \emph{IEEE Transactions on Transportation Electrification}, vol.~7, no.~3, pp. 2010--2022, 2020.

\bibitem{babu2024regression}
M.~S. Babu, Y.~Tiwari, M.~Pal \emph{et~al.}, ``Regression based anomaly detection in electric vehicle state of charge fluctuations through analysis of evci data,'' \emph{arXiv preprint arXiv:2401.01580}, 2024.

\bibitem{wang2021data}
Z.~Wang, C.~Song, L.~Zhang, Y.~Zhao, P.~Liu, and D.~G. Dorrell, ``A data-driven method for battery charging capacity abnormality diagnosis in electric vehicle applications,'' \emph{IEEE Transactions on Transportation Electrification}, vol.~8, no.~1, pp. 990--999, 2021.

\bibitem{archive}
S.~Bai, J.~Z. Kolter, and V.~Koltun, ``An empirical evaluation of generic convolutional and recurrent networks for sequence modeling,'' \emph{arXiv preprint arXiv:1803.01271}, 2018.

\bibitem{nair2010rectified}
V.~Nair and G.~E. Hinton, ``Rectified linear units improve restricted boltzmann machines,'' in \emph{Proceedings of the 27th international conference on machine learning (ICML-10)}, 2010, pp. 807--814.

\bibitem{salimans2016weight}
T.~Salimans and D.~P. Kingma, ``Weight normalization: A simple reparameterization to accelerate training of deep neural networks,'' \emph{Advances in neural information processing systems}, vol.~29, 2016.

\bibitem{srivastava2014dropout}
N.~Srivastava, G.~Hinton, A.~Krizhevsky, I.~Sutskever, and R.~Salakhutdinov, ``Dropout: a simple way to prevent neural networks from overfitting,'' \emph{The journal of machine learning research}, vol.~15, no.~1, pp. 1929--1958, 2014.

\bibitem{conference}
M.~Thill, W.~Konen, and T.~B{\"a}ck, ``Time series encodings with temporal convolutional networks,'' in \emph{International Conference on Bioinspired Methods and Their Applications}.\hskip 1em plus 0.5em minus 0.4em\relax Springer, 2020, pp. 161--173.

\bibitem{elsiever}
M.~Thill, W.~Konen, H.~Wang, and T.~B{\"a}ck, ``Temporal convolutional autoencoder for unsupervised anomaly detection in time series,'' \emph{Applied Soft Computing}, vol. 112, p. 107751, 2021.

\bibitem{de2000mahalanobis}
R.~De~Maesschalck, D.~Jouan-Rimbaud, and D.~L. Massart, ``The mahalanobis distance,'' \emph{Chemometrics and intelligent laboratory systems}, vol.~50, no.~1, pp. 1--18, 2000.

\bibitem{chargestd}
``E- amrit,'' \url{https://e-amrit.niti.gov.in/standards-and-specifications}.

\bibitem{wu2022hierarchical}
Y.~Wu, Z.~Wang, Y.~Huangfu, A.~Ravey, D.~Chrenko, and F.~Gao, ``Hierarchical operation of electric vehicle charging station in smart grid integration applications—an overview,'' \emph{International Journal of Electrical Power \& Energy Systems}, vol. 139, p. 108005, 2022.

\bibitem{chen2020review}
T.~Chen, X.-P. Zhang, J.~Wang, J.~Li, C.~Wu, M.~Hu, and H.~Bian, ``A review on electric vehicle charging infrastructure development in the uk,'' \emph{Journal of Modern Power Systems and Clean Energy}, vol.~8, no.~2, pp. 193--205, 2020.

\bibitem{zhu2016influence}
C.~Zhu and D.~Gao, ``Influence of data preprocessing,'' \emph{Journal of Computing Science and Engineering}, vol.~10, no.~2, pp. 51--57, 2016.

\bibitem{kwon2018empirical}
D.~Kwon, K.~Natarajan, S.~C. Suh, H.~Kim, and J.~Kim, ``An empirical study on network anomaly detection using convolutional neural networks,'' in \emph{2018 IEEE 38th International Conference on Distributed Computing Systems (ICDCS)}.\hskip 1em plus 0.5em minus 0.4em\relax IEEE, 2018, pp. 1595--1598.

\bibitem{choi2022multivariate}
T.~Choi, D.~Lee, Y.~Jung, and H.-J. Choi, ``Multivariate time-series anomaly detection using seqvae-cnn hybrid model,'' in \emph{2022 International Conference on Information Networking (ICOIN)}.\hskip 1em plus 0.5em minus 0.4em\relax IEEE, 2022, pp. 250--253.

\bibitem{zhao2020multivariate}
H.~Zhao, Y.~Wang, J.~Duan, C.~Huang, D.~Cao, Y.~Tong, B.~Xu, J.~Bai, J.~Tong, and Q.~Zhang, ``Multivariate time-series anomaly detection via graph attention network,'' in \emph{2020 IEEE International Conference on Data Mining (ICDM)}.\hskip 1em plus 0.5em minus 0.4em\relax IEEE, 2020, pp. 841--850.

\end{thebibliography}

\end{document}